\documentclass[10pt,letterpaper,oneside,onecolumn]{article}
\usepackage[]{fontenc}
\usepackage[dvips]{epsfig}

\makeatletter
\def\LyX{L\kern-.1667em\lower.25em\hbox{Y}\kern-.125emX\spacefactor1000}%
\newcommand{\lyxtitle}[1] {\thispagestyle{empty}
\global\@topnum\z@
\section*{\LARGE \centering \sffamily \bfseries \protect#1 }
}

\newcommand{\lyxletterstyle}{
\setlength\parskip{0.7em}
\setlength\parindent{0pt}
}

\makeatother

\lyxletterstyle
\begin{document}

{\bfseries \large \hfill{}\normalsize Disentanglement of pure
 bipartite quantum states by local cloning\hfill{}\large  \par}

\( \smallskip  \)

{\large \hfill{}\normalsize Somshubhro Bandyopadhyay\( ^{a,} \)\footnote{
email: dhom@boseinst.ernet.in
} , Guruprasad Kar\( ^{b} \) and Anirban Roy\( ^{b} \) \hfill{} \large \par}

\( \smallskip  \)

{\em \( ^{a} \)\em Department of Physics, Bose Institute, 93/1 A.P.C. Road,
Calcutta -700009, India\hfill{}\par}

{\em \( ^{b} \)\em Physics and Applied Mathematics Unit, Indian Statistical
Institute, 203 B.T. Road, Calcutta -700035, India\par}

\( \smallskip  \)

\bfseries \hfill{}Corresponding author\hfill{} \mdseries 

\hfill{}Somshubhro Bandyopadhyay\hfill{} 

\hfill{}Department of Physics\hfill{} 

\hfill{}Bose Institute\hfill{} 

\hfill{}93/1 A.P.C. Road\hfill{} 

\hfill{}Calcutta - 700009\hfill{} 

\hfill{}INDIA\hfill{} 

email: dhom@boseinst.ernet.in

Telephone: 91 - 33 - 350 - 2402/03, 91 - 33 - 350-6702

FAX: 91 - 33 - 350 - 6790

\( \smallskip  \)

{\bfseries \hfill{}Abstract\hfill{} \par}

{We discuss disentanglement of pure bipartite quantum states within
the framework of the schemes developed for entanglement splitting
and broadcasting of entanglement.\par}

\( \pagebreak  \)

{\bfseries \large 1. Introduction\par}

A composite quantum system consisting of two subsystems is said to
be entangled or inseparable if in general the density matrix cannot
be written as 

\hfill{}\( \rho =\sum w_{i}\rho _{i}^{(1)}\otimes \rho _{i}^{(2)} \)\hfill{}(1)

where the positive weights \( w_{i} \) satisfy \( \sum w_{i}=1 \) and \( \rho ^{(1)}=Tr_{_{2}}(\rho ) \) and \( \rho ^{(2)}=Tr_{_{1}}(\rho ) \) are the reduced
density matrices of the subsystems. Therefore the density matrix of
a separable or disentangled system can always be written in the form
(1). So the question is whether there exisits any universal transformation
acting on any given arbitrary entangled state \( \rho  \) transforms it into
a state that can be written in the form (1). If really such transformation
exists then we have an ideal disentanglement machine \( (DM) \). Thus disentanglement
is the process which transforms an inseparable or entangled state
(consisting of two qubits) into a separable one such that the reduced
density operators of the individual subsystems remain unchanged. 

First we give two definitions of disentanglement [1, 2]:

\em Definition 1 \em - Disentanglement is the process that transforms
a state of two (or more) subsystems into an unentangled state (in
general, a mixture of product states) such that the reduced density
matrices of each of the subsystems are unaffaected.

\em Definition 2 \em - Disentanglement into a tensor product state
is the process that transforms a state of two (or more) subsystems
into a tensor product of the two reduced density matrices.

Note that the second definition is a special case of the first one.

Recently Terno [2] showed that universal disentanglement into product
states is impossible. Tal Mor [1] investigated the possibility of
universal disentanglement into separable states where he showed that
a universal disentangling machine cannot exist. We now briefly describe
what the above stated two results actually mean.

The first result (Terno[2]) tells us that

\hfill{}\( \rho ^{ent}\; \; \frac{DM}{}\triangleright \; \; \rho ^{(1)}\otimes \rho ^{(2)} \)\hfill{} (2)

is not allowed.

The second result (T. Mor [1]) negates the possibilty of

\hfill{}\( \rho ^{ent}\; \; \frac{DM}{}\triangleright \; \; \rho ^{disent} \)\hfill{}(3) 

such that \hfill{}\( \rho ^{(i)}=Tr_{j}(\rho ^{ent})=Tr_{j}(\rho ^{disent});\; \; i\neq j;\; \; i,j=1,2 \)\hfill{} 

So the question is then how well we can disentangle a pure bipartite
quantum state ? The answer lies in the possibilty of constructing
an \( UDM \) such that the reduced density operators of \( \rho ^{disent} \) are \em close \em to
the reduced density operators corresponding to \( \rho ^{ent} \). What is partcularly
meant by closeness will be clear from the following discussion. We
require that the reduced state operators of the subsystems after disentanglement
should remain in a scaled form (isotropy requirement) i.e. \( \rho _{ad}^{(i)}=\eta \rho _{bd}^{(i)}+(\frac{1-\eta }{2})I \)\( \; \;  \)\footnote{
\( ad \) and \( bd \) stands for after disentanglement and before disentanglement
respectively
}, then \( \eta  \) being the scaling parameter (can take values between 0 and
1 in principle) stands as a measure of closeness. For example \( \eta =1 \) tells
us that the reduced state operator after disentanglement hasn't changed
whereas \( \eta =0 \) gives the information that the subsystems have become totally
random after disentanglement etc. 

Keeping in mind the above constraints we discuss how the schemes developed
for entanglement splitting [3] and broadcasting of entanglement [4,
5] can be used for disentangling a pure entangled state. 

\( \smallskip  \)

{\bfseries \large 2. Disentanglement of a two qubit pure entangled
 state\par}

In what follows we discuss how disentanglement of a pure two qubit
entangled state can be achieved by local cloning of the qubits within
the existing schemes of entanglement splitting and broadcasting of
entanglement. Before that we recall what precisely we want to achieve.
Our aim is to have 

\hfill{}\( \rho ^{ent}\; \; \frac{DM}{}\triangleright \; \; \rho ^{disent} \)\hfill{}(3a) 

such that \hfill{}\( \rho _{bd}^{(i)}=Tr_{j}(\rho ^{ent})\: is\: close\: to\: \rho _{ad}^{(i)}=Tr_{j}(\rho ^{disent});\; \; i\neq j;\; \; i,j=1,2 \).\hfill{} 

In particular we want  

\hfill{}\( \rho _{ad}^{(i)}=\eta \rho _{bd}^{(i)}+\left( \frac{1-\eta }{2}\right) I \)\hfill{}(3b) 

such that \( \eta  \) takes the maximum possible value.

{\bfseries 2. 1. Disentanglement by local cloning of one qubit\par}

First we consider how we can disentangle a two qubit pure entangled
state by local cloning of any one qubit. The local cloning of a single
qubit of a two particle entangled state forms the basis of entanglement
splitting [3] that can be defined as the process by which any one
of a two party entangled system transfers a part of his entanglement
to a third party. Suppose we have two parties \bfseries \( x \) \mdseries and
\( y \) sharing an entangled state of two qubits given by

\hfill{}\( \left| \psi \right\rangle =\alpha \left| 00\right\rangle _{xy}+\beta \left| 11\right\rangle _{xy} \)\hfill{}.

The first qubit belongs to \( x \) and the second belongs to \( y \) as usual.
Now the qubit belonging to any one (say \bfseries \( x \)\mdseries )
of the two is cloned and let the two copies be \( x_{1} \) and \( x_{2} \). This gives
rise to a composite system \( \rho _{x_{1}x_{2}y} \) consisting of three qubits. Tracing out
one of the copies \( x_{i}(i=1,2) \) produces a two qubit composite system which is
inseparable under certain conditions. This is the basic concept of
entanglement splitting. Note that in this process nothing is done
to the qubit that belongs to \( y \). Thus the reduced density matrix corresponding
to \( y \) remains unchanged.

Consider the following universal cloning transformation for local
copying of the subsystem \( x \), defined by

\hfill{}\( U\left| 0\right\rangle \left| \right\rangle \left| Q\right\rangle =a\left| 00\right\rangle \left| A\right\rangle +b(\left| 01\right\rangle +\left| 10\right\rangle )\left| B\right\rangle  \)\hfill{} (4)

\hfill{}\( U\left| 1\right\rangle \left| \right\rangle \left| Q\right\rangle =a\left| 11\right\rangle \widetilde{\left| A\right\rangle }+b(\left| 01\right\rangle +\left| 10\right\rangle )\widetilde{\left| B\right\rangle } \)\hfill{} (5)

where \( \left| \right\rangle  \)denotes the blank qubit supplied to the cloner, \( \left| Q\right\rangle  \) denotes the
initial state of the quantum copier (ancilla), \( \left| A\right\rangle ,\left| B\right\rangle ,\widetilde{\left| A\right\rangle },\widetilde{\left| B\right\rangle } \) are the normalized
ancilla output states. The coefficients \( a \) and \( b \) are in general complex.
The following conditions hold from unitarity, isotropy and symmetry
requirements for an universal quantum cloner [6]

\hfill{} \( \left| a\right| ^{2}+2\left| b\right| ^{2}=1 \)\hfill{} (6)

\hfill{}\( \left\langle B\right| \widetilde{\left. B\right\rangle }=\left\langle A\mid B\right\rangle =\widetilde{\left\langle A\right. }\widetilde{\left| B\right\rangle }=0 \)\hfill{} (7)

The fidelity of the above universal quantum cloner defined by the
transformations \( (4) \) and \( (5) \) along with the conditions \( (6) \) and \( (7) \), is given
by

\hfill{}\( F=\frac{1}{2}(1+\eta ) \)\hfill{} (8)

where the reduction factor (also known as the Black Cow factor [7])
\( \eta  \) is given by [6]

\hfill{}\( \eta =\left| a\right| ^{2}=Re\left( ab^{*}\widetilde{\left\langle B\right| }\left. A\right\rangle +a^{*}b\widetilde{\left\langle A\right| }\left. B\right\rangle \right)  \)\hfill{} (8)

Choosing, \( \widetilde{\left\langle B\right. }\left| A\right\rangle =\widetilde{\left\langle A\right| }\left. B\right\rangle =1 \) one obtains the optimal quantum cloner [8,9] for which
\( \eta =2/3 \). Thus a less optimal quantum cloner but nevertheless universal (isotropic)
can be constructed by varying the scalar product of the ancilla output
states. 

We now apply this cloning transformation to copy the subsystem \( x \) to
produce the copies \( x_{1} \) and \( x_{2} \). After tracing out the cloning machine
part and any one of the copies the resulting density matrix is given
by

\( \rho _{x_{i}y}=\left( \frac{1+\eta }{2}\right) \left( \alpha ^{2}\left| 00\right\rangle \left\langle 00\right| +\beta ^{2}\left| 11\right\rangle \left\langle 11\right| \right) +\alpha \beta \eta \left( \left| 00\right\rangle \left\langle 11\right| +\left| 11\right\rangle \left\langle 00\right| \right)  \)

\hfill{}\( +\left( \frac{1-\eta }{2}\right) \left( \alpha ^{2}\left| 01\right\rangle \left\langle 01\right| +\beta ^{2}\left| 10\right\rangle \left\langle 10\right| \right)  \)\hfill{} \( i=1,2 \)\hfill{} (9)

Applying the Peres - Horodecki theorem [10,11] to test the inseparability
of \( \rho _{x_{i}y} \) it turns out that the state is inseparable for all values of
\( \alpha  \) provided \( \eta >1/3 \) (Note that upper bound of \( \eta  \) is 2/3). Thus it is possible
to achieve disentanglement of any arbitrary pure two particle entangled
state provided we employ an universal (isotropic) cloner whose fidelity
\( F\leq 2/3 \) (recall that \( \eta  \) and \( F \) are related by Eq. 8) to copy one of the qubits.
Now that our requirement is also to have reduced density matrices
of the disentangled state as close as possible to those of the entangled
one, we note that the reduced density matrix of the subsystem \( y \) is
unaltered whereas that of the subsystem \( x \) is changed. It is clear
that the subsystem \( x \) is copied rather poorly since the maximum fidelity
with which it has to be copied to achieve disentanglement is 2/3.
Thus after disentanglement by this process, although one suceeds in
keeping one subsystem unchanged but ends up with a rather poor copy
of the other. Let us summarise the results of the preceeding section.

1. It is possible to disentangle any arbitrary bipartite entangled
state by applying local cloning on one of its qubits provided the
reduction factor of the isotropic cloner is less than or equal to
\( 1/3 \) (i.e. fidelity \( F\leq 2/3 \)).

2. After disentanglement the reduced density matices of the subsystems
are given by

\hfill{}\( \rho _{ad}^{(y)}=\rho _{bd}^{(y)} \)\hfill{} 

\hfill{}\( \rho _{ad}^{(x)}=\eta \rho _{bd}^{(x)}+\left( \frac{1-\eta }{2}\right) I \) \hfill{}where \( \eta _{max}=1/3 \). 

{\bfseries 2. 2. Disentanglement by local cloning of both the qubits
 \par}

In the previous section we showed how to disentangle pure states by
applying local cloning operation on one of the qubits. In this section
we use the concept of broadcasting quantum inseparability  via local
copying, first shown to be possible by Buzek et al [4]. First we briefly
sketch the essentials of broadcasting of entanglement where the entanglement
originally shared by a single pair is transferred into two less entangled
pairs using only local operations. Suppose two distant parties \( a_{1} \) and
\( a_{2} \) share an entangled two qubit state

\hfill{}\( \left| \psi \right\rangle =\alpha \left| 00\right\rangle _{a_{1}a_{2}}+\beta \left| 11\right\rangle _{a_{1}a_{2}} \)\hfill{} \( \left( 10\right)  \)

where \( \alpha ^{2}+\beta ^{2}=1 \) and \( \alpha ,\beta  \) are real. 

The first qubit belongs to \( a_{1} \) and the second belongs to \( a_{2} \). Each of
the two parties now performs local cloning operations on their own
qubit. It turns out that for some values of \( \alpha  \),

\( \left( a\right)  \) non local output states are inseparable, and

\( \left( b\right)  \) local output states are separable

hold simultaneously. Buzek et al. [4] used optimal quantum cloners
for local copying of the subsystems and showed that the nonlocal outputs
are inseparable if 

\hfill{}\( \frac{1}{2}-\frac{\sqrt{39}}{16}\leq \alpha ^{2}\leq \frac{1}{2}+\frac{\sqrt{39}}{16} \) \hfill{}(11) 

Now consider the same system defined by (10). The two subsystems \( a_{i}(i=1,2) \)
are locally copied according to the cloning transformations (4) and
(5) to produce output two systems \( b_{i}(i=1,2) \). The local output state of a copier
is given by the density operator

\hfill{}\( \widehat{\rho }_{a_{i}b_{i}}^{(out)} \)\( =\alpha ^{2}\eta \left| 00\right\rangle \left\langle 00\right| +\beta ^{2}\eta \mid \left| 11\right\rangle \left\langle 11\right| +(1-\eta )\left| +\right\rangle \left\langle +\right|  \)\footnote{
\( \left| +\right\rangle =\frac{1}{\sqrt{2}}\left( \left| 01\right\rangle +\left| 10\right\rangle \right)  \)
}\hfill{} (12)

and the nonlocal output is described by the density operator

\hfill{}\( \widehat{\rho }_{a_{i}b_{j}}^{(out)}=[\alpha ^{2}\eta +(\frac{1-\eta }{2})^{2}]\left| 00\right\rangle \left\langle 00\right| +[\beta ^{2}\eta +(\frac{1-\eta }{2})^{2}]\left| 11\right\rangle \left\langle 11\right|  \)\hfill{} 

\hfill{}\hfill{}\hfill{}\hfill{}\hfill{}\( +(\frac{1-\eta ^{2}}{4})(\left| 01\right\rangle \left\langle 01\right| +\left| 10\right\rangle \left\langle 10\right| )+\alpha \beta \eta ^{2}(\left| 00\right\rangle \left\langle 11\right| +\left| 11\right\rangle \left\langle 00\right| ) \)\hfill{} 

\hfill{}\( i\neq j;\; \; \; i,j=1,2 \)\( \; \; \; \; \; \; \;  \)(13)

It follows from the Peres-Horodecki theorem [10, 11] that \( \widehat{\rho }_{a_{i}b_{j}}^{(out)} \) is inseparable
if 

\hfill{}\( \frac{1}{2}-[\frac{1}{4}-\frac{(1-\eta ^{2})^{2}}{16\eta ^{4}}]^{1/2}\leq \alpha ^{2}\leq \frac{1}{2}+[\frac{1}{4}-\frac{(1-\eta ^{2})^{2}}{16\eta ^{4}}]^{1/2} \)\hfill{}(14) 

The requirement that \( [\frac{1}{4}-\frac{\xi ^{2}(1-\xi )^{2}}{(1-2\xi )^{4}}]^{1/2} \) has to be positive otherwise the domain of
\( \alpha ^{2} \) would be meaningless leads to the lower bound of \( \eta  \),

\hfill{}\( \eta \geq \sqrt{\frac{1}{3}} \)\hfill{} (15)

The upper bound is of course \( 2/3 \) corresponding to the optimal quantum
cloner. 

Again applying the Peres-Horodecki theorem it is easy to obtain that
\( \widehat{\rho }_{a_{i}b_{i}}^{(out)} \) is separable if

\hfill{}\( \frac{1}{2}-\{\frac{1}{4}-\frac{(1-\eta )^{2}}{4\eta ^{2}}\}^{1/2}\leq \alpha ^{2}\leq \frac{1}{2}+\{\frac{1}{4}-\frac{(1-\eta )^{2}}{4\eta ^{2}}\}^{1/2} \)\hfill{}(16) 

As one can observe comparing (14) and (16) that \( \widehat{\rho }_{a_{i}b_{i}}^{(out)} \) is separable if
\( \widehat{\rho }_{a_{i}b_{j}}^{(out)} \) is inseparable. 

Here our interest lies in the possibilty of disentangling the state
(10) and therefore we note from inequality (15) that the state (13)
becomes separable for all \( \alpha  \) when \( \eta <1/\sqrt{3} \) . But when \( \frac{1}{\sqrt{3}}\leq \eta \leq \frac{2}{3} \) its still possible
to disentangle the state (10) but not for all \( \alpha  \) as is clear from (14).
Besides our objective is also to copy the subsystem in the best possible
way for which one has to use optimal quantum cloners for local copying.
In that case if and only if \( \alpha  \) lies outside the range given by (11)
the final state becomes disentangled with the best possible reduced
density matrices of the subsystems. Now let us summarise the main
results ,

1. An arbitrary bipartite pure entangled state can be disentangled
by local cloning of the individual subsystems provided the reduction
factor of the isotropic cloners used is less than \( 1/\sqrt{3} \)(ie. fidelity \( F<\frac{\sqrt{3}+1}{2\sqrt{3}} \)).

2. After disentanglement of the original state the reduced density
matrices of the subsystems are given by

\hfill{}\( \rho _{ad}^{\left( a_{i}\right) }=\eta \rho _{bd}^{\left( a_{i}\right) }+\left( \frac{1-\eta }{2}\right) I \)\hfill{} \( i=1,2 \)

where \( \eta <1/\sqrt{3} \).

Its clear from the above discussions that its impossible to disentangle
an arbitrary pure two particle entangled state by using optimal quantum
cloners to copy the subsystems locally such that qualitatively best
possible reduced density operators of the subsystems can be obtained.

Here we observe that all the above discussions particularly hold good
when we use a isotropic quantum cloner of \( 1\rightarrow 2 \) type. So we can as well
look for the possibility of disentanglement using an optimal cloner
of type \( 1\rightarrow M \) for some \( M(M>2) \) to copy the qubits locally. Since we know that
quality of copies decrease with increasing \( M \), therefore we should
look for the smallest value of \( M(M>2) \) such that applying an optimal \( 1\rightarrow M \) quantum
cloner to copy the qubits locally we will be able to disentangle a
pure state. It has already been shown [5] that producing three copies
of any pure entanglement by applying \( 1\rightarrow 3 \) optimal quantum cloner for
local copying of the individual subsystems is forbidden. So the two
criteria are satisfied for \( M=3 \). The important thing to note in this
case is the fidelity of an optimal \( 1\rightarrow 3 \) quantum cloner which is 7/9.
We have seen earlier that disentanglement of any pure state can be
achieved if we clone the qubits locally by an \( 1\rightarrow 2 \) isotropic cloner of
fidelity \( F<\frac{\sqrt{3}+1}{2\sqrt{3}} \)(see Eq. (8) and the discussion after inequality (16)) whereas
the fidelity of each local qubit in the later case being 7/9. Although
disentanglement is achieved in both the cases for any arbitrary pure
entangled state the previous one is better since \( \frac{\sqrt{3}+1}{2\sqrt{3}}>7/9 \) implying that locally
individual systems are better copied.

\( \smallskip  \)

Till now we have described how one is able to disentangle a pure two
qubit entangled state working within the existing schemes of entanglement
splitting and broadcasting of entanglement. One distinct advantage
of the first scheme (where we have made use of the concept of entanglement
splitting) is that the state of one of the subsystems remains unchanged
though the copy of the other subsystem is rather poor \( [\eta _{threshold}=1/3;Fidelity_{threshold}=\frac{1}{2}(1+\eta )=2/3=0.666] \) as compared
to the second where both the subsystems undergo change in their respective
states but the copies being better \( (\eta _{threshold}<1/\sqrt{3};Fidelity_{threshold}<\frac{1}{2}(1+\eta )=\frac{\sqrt{3}+1}{2\sqrt{3}}=0.788) \). Nevertheless the second scheme
allows the possibilty of using the optimal cloner provided the parameter
\( \alpha  \) of the original entangled state lies outside the range specified
by (11). So which scheme is better  can probably be best justified
from the motivation of a given problem where disentanglement is required.

\( \smallskip  \)

{\bfseries \large 3. Conclusion\par}

Since universal disentanglement into separable states is not allowed
we have explored how well one can approximate universal disentanglement
of a pure bipartite quantum state. In other words we have tried to
answer possibly ``how good'' an universal disentanglement machine can
be since construction of a perfect universal disentanglement machine
is forbidden. We have shown that working with less optimal but isotropic
quantum cloner within the framework of entanglement splitting and
broadcasting of entanglement it is possible to disentangle a pure
bipartirte quantum state. 

\( \medskip  \)

{\bfseries \large References\par}

[1] T. Mor, quant-ph/9812020.

[2] D. Terno, quant-ph/9811036.

[3] D. Bru\( \ss  \), quant-ph/9902023.

[4] V. Buzek, V. Vedral, M. B. Plenio, P. L. Knight and M. Hillery,
Phys. Rev. A 55 (1997) 3327.

[5] S. Bandyopadhyay and G. Kar, quant-ph/9902073.

[6] D. Bru\( \ss  \), D. P. DiVincenzo, A. Ekert, C. A. Fuchs, C. Macchiavello
and J. A. Smolin, Phys. Rev. A 57 (1998) 2368.

[7] R. Werner, quant-ph/9804001.

[8] V. Buzek and M. Hillery, Phys. Rev. A 54 (1996) 1844.

[9] N. Gisin and S. Massar, Phys. Rev. Lett. 79 (1997) 2153.

[10] A. Peres, Phys. Rev. Lett. 77 (1996) 1413.

[11] M. Horodecki, P. Horodecki and R. Horodecki, Phys. Lett. A 223
(1996) 1.

\end{document}